\documentclass[a4paper,12pt]{article}
\usepackage{graphicx,rotating}
\usepackage{hyperref}\usepackage{slashed}
\pdfoutput=1
\usepackage{charter}

\hypersetup{colorlinks,bookmarksopen,bookmarksnumbered,
linkcolor=blus,pdfstartview=FitH,urlcolor=rossos,citecolor=verde}

\newcommand{\fig}[1]{~\ref{fig:#1}}
\newcommand{\gs}{f_0}
\newcommand{\gt}{f_2}

\usepackage{multicol}
\usepackage{color}
\definecolor{rosso}{cmyk}{0,1,1,0.4}
\definecolor{rossos}{cmyk}{0,1,1,0.55}
\definecolor{rossoc}{cmyk}{0,1,1,0.2}
\definecolor{blu}{cmyk}{1,1,0,0.3}
\definecolor{blus}{cmyk}{1,1,0,0.6}
\definecolor{bluc}{cmyk}{1,1,0,0.1}
\definecolor{verde}{cmyk}{0.92,0,0.59,0.25}
\definecolor{verdec}{cmyk}{0.92,0,0.59,0.15}
\definecolor{verdes}{cmyk}{0.92,0,0.59,0.4}

\newcommand{\riga}[1]{\noalign{\hbox{\parbox{\textwidth}{#1}}}\nonumber}

\newcommand{\mub}{\bar{\mu}}

\newcommand{\eq}[1]{~{\rm (\ref{eq:#1})}}

\newcommand{\GeV}{\,{\rm GeV}}

\newcommand{\Tr}{\,{\rm Tr}}
\newcommand{\diag}{\,{\rm diag}}

\def\circa#1{\,\raise.3ex\hbox{$#1$\kern-.75em\lower1ex\hbox{$\sim$}}\,}

\newcommand{\beq}{\begin{equation}}
\newcommand{\eeq}{\end{equation}}

\newcommand{\bea}{\begin{eqnarray}}
\newcommand{\eea}{\end{eqnarray}}
\newcommand{\be}{\begin{equation}}
\newcommand{\ee}{\end{equation}}
\font\tenrsfs=rsfs10 at 12pt
\font\sevenrsfs=rsfs7
\font\fiversfs=rsfs5
\newfam\rsfsfam
\textfont\rsfsfam=\tenrsfs
\scriptfont\rsfsfam=\sevenrsfs
\scriptscriptfont\rsfsfam=\fiversfs
\def\mathscr#1{{\fam\rsfsfam\relax#1}}

\def\Lag{\mathscr{L}}

\def\circa#1{\,\raise.3ex\hbox{$#1$\kern-.75em\lower1ex\hbox{$\sim$}}\,}
\makeatletter

\def\hhref#1{\href{http://arxiv.org/abs/#1}{arXiv:#1}} % in bibliography

\def\art{\@ifnextchar[{\eart}{\oart}}
\def\eart[#1]#2#3#4#5#6{{\rm #2}, {\em #3 \bf #4} {\rm (#6) #5} ({\em #1})}
\def\hepart[#1]#2{{\rm #2, \hhref{#1}}}
\newcommand{\oart}[5]{{\rm #1}, {\em #2 \bf #3} {\rm (#5) #4}}

\newcounter{alphaequation}[equation]
\def\thealphaequation{\theequation\hbox to
0.6em{\hfil\alph{alphaequation}\hfil}}
\def\eqnsystem#1{
\def\@eqnnum{{\rm (\thealphaequation)}}
\def\@@eqncr{\let\@tempa\relax \ifcase\@eqcnt \def\@tempa{& & &} \or
  \def\@tempa{& &}\or \def\@tempa{&}\fi\@tempa
  \if@eqnsw\@eqnnum\refstepcounter{alphaequation}\fi
\global\@eqnswtrue\global\@eqcnt=0\cr}
\refstepcounter{equation} \let\@currentlabel\theequation \def\@tempb{#1}
\ifx\@tempb\empty\else\label{#1}\fi
\refstepcounter{alphaequation}
\let\@currentlabel\thealphaequation
\global\@eqnswtrue\global\@eqcnt=0 \tabskip\@centering\let\\=\@eqncr
$$\halign to \displaywidth\bgroup \@eqnsel\hskip\@centering
$\displaystyle\tabskip\z@{##}$&\global\@eqcnt\@ne
\hskip2\arraycolsep\hfil${##}$\hfil& \global\@eqcnt\tw@\hskip2\arraycolsep
$\displaystyle\tabskip\z@{##}$\hfil
\tabskip\@centering&\llap{##}\tabskip\z@\cr}
\def\endeqnsystem{\@@eqncr\egroup$$\global\@ignoretrue} \makeatother

\newcommand{\eV}{\,{\rm eV}}

\begin{document}

\vspace{1cm}

\begin{center}
{\LARGE \bf \color{rossos}
Agravity}\\[1cm]

{\large\bf Alberto Salvio$^{a}$ {\rm and} Alessandro Strumia$^{b}$}  
\\[7mm]
{\it $^a$ } {\em Departamento de F\'isica Te\'orica, Universidad Aut\'onoma de Madrid\\ and Instituto de F\'isica Te\'orica IFT-UAM/CSIC,  Madrid, Spain}\\[3mm]
{\it $^b$ Dipartimento di Fisica dell'Universit{\`a} di Pisa and INFN, Italy\\
and National Institute of Chemical Physics and Biophysics, Tallinn, Estonia}

\vspace{1cm}
{\large\bf\color{blus} Abstract}
\begin{quote}
We explore the possibility that the fundamental theory of nature does not contain any scale.
This implies a renormalizable quantum gravity theory where 
the graviton kinetic term has 4 derivatives, 
and can be reinterpreted as gravity minus an anti-graviton.
We compute the super-Planckian RGE of  adimensional gravity coupled to a generic matter sector.
The Planck scale and a flat space can arise dynamically at quantum level
provided that a quartic scalar coupling and its $\beta$ function vanish at the Planck scale.
This is how the Higgs boson behaves for $M_h\approx 125\GeV$ and $M_t\approx 171\GeV$.
Within agravity, inflation is a generic phenomenon: the slow-roll parameters are given by the $\beta$-functions of the theory,
and are small if couplings are perturbative.
The predictions $n_s\approx 0.967$ and $r\approx 0.13$ arise if the inflaton is identified with the Higgs of gravity.
Furthermore, quadratically divergent corrections to the Higgs mass vanish:
a small weak scale is natural and can be generated by agravity quantum corrections.

\end{quote}

\thispagestyle{empty}
\end{center}
\begin{quote}
{\large\noindent\color{blus} 
}

\end{quote}
\vspace{-1.5cm}

\tableofcontents

\setcounter{footnote}{0}

\section{Introduction}
We propose a general principle that leads to a renormalizable and predictive
theory of quantum gravity
where all scales are generated dynamically,
where a small weak scale coexists with the Planck scale,
where inflation is a natural phenomenon.
The price to pay is a ghost-like anti-graviton state.

The general principle is:
{\em nature does not possess any scale}. 
We start presenting how this principle is suggested by two recent experimental results,
and next discuss its implementation and  consequences.

\subsubsection*{1) Naturalness}
In the past decades theorists assumed that Lagrangian
terms with positive mass dimension
(the Higgs mass $M_h$ and the vacuum energy) receive
big power-divergent quantum corrections, as suggested by Wilsonian computation techniques
that attribute physical meaning to momentum shells of loop integrals~\cite{Wilson}.
According to this point of view, a modification of the SM at the weak scale is needed to
make quadratically divergent corrections to $M_h^2$ naturally small.  Supersymmetry seems the most successful possibility, but naturalness got increasingly challenged by
the non-observation of any new physics that keeps the weak scale naturally small~\cite{anat}.

The naturalness problem can be more generically formulated as a problem of 
 the effective theory ideology, according to which  
nature is described by a non-renormalizable Lagrangian of the form
\beq \Lag \sim \Lambda^4 + \Lambda^2  |H|^2  +\lambda |H|^4 + \frac{|H|^6}{\Lambda^2}+\cdots \eeq
where, for simplicity, we wrote only the Higgs potential terms.
The assumption that $\Lambda\gg M_h$ explains why at low energy $E\sim M_h$ we only observe
those terms not  suppressed by $\Lambda$: the renormalizable interactions.
Conservation of baryon number, lepton number, and other successful features of the Standard Model
indicate a large  $\Lambda\circa{>}10^{16}\GeV$.
In this context, gravity can be seen as a non-renormalizable interaction suppressed by $\Lambda \sim M_{\rm Pl}=1.22~10^{19}\GeV$.

\smallskip

However, this scenario also leads to the expectation that particles cannot be light unless protected by a symmetry.
The Higgs mass should be $M_h^2\sim \Lambda^2$ and the vacuum energy should be $V\sim \Lambda^4$.
In nature, they are many orders of magnitude smaller, and no protection mechanism is observed so far.

We assume that this will remain the final experimental verdict and try to derive the theoretical implications.

Nature is maybe telling us that both super-renormalizable terms and non-renormalizable terms vanish and that
only adimensional interactions exists.

\subsubsection*{2) Inflation}
Cosmological observations suggest inflation with a small amount of anisotropies.
However, this is a quite unusual outcome of quantum field theory:
it requires special models with flat potentials, and often field values above the Planck scale.
Let us discuss this issue in the context of 
Starobinsky-like inflation models~\cite{xii}: a class of inflation models favoured by Planck data~\cite{Planck}.
%Such models can be presented in different equivalent languages.
Such models can be described in terms of one scalar $S$ (possibly identified with the Higgs $H$)
with a potential $V(S)$ and a coupling to gravity $-\frac12 f(S) R$.
Going to the Einstein frame (i.e.\ making field redefinitions such that the graviton kinetic term $R$
gets its canonical coefficient) the potential gets rescaled into $V_E=\bar M_{\rm Pl}^4 V/f^2$,
where $\bar M_{\rm Pl}=M_{\rm Pl}/\sqrt{8\pi} =2.4~10^{18}\GeV$ is the reduced Planck mass.
Special assumptions such as $V(S) \propto f(S)^2$ make the Einstein-frame
potential $V_E$ flat at $S\gg M_{\rm Pl}$, with predictions compatible with present observations~\cite{xii}.
However this flattening is the result of a fine-tuning: in presence of generic Planck-suppressed operators
$V$ and $f$ and thereby $V_E$ are generic functions of $S/M_{\rm Pl}$.
 
Nature is maybe telling us that $V_E$ becomes flat at $S\gg M_{\rm Pl}$ because only adimensional terms exists.

\subsubsection*{The principle}
These observations vaguely indicate that nature prefers adimensional terms, so that
ideas along these lines are being discussed in the literature~\cite{FNmodels}.

%There are different options for converting  this vague idea into a concrete theory.

\bigskip

We propose a simple concrete principle: {\em the fundamental theory of nature does not possess any mass or length scale}
and thereby only contains  `renormalizable' interactions --- i.e.\ interactions with dimensionless couplings.
%\footnote{ from
%the string multiverse scenario, where the Planck mass is related to the string length,
%where the complexity of the SM Lagrangian comes from compactification over a complicated extra-dimensional background, where the huge number of possible vacua is used to
%explain the smallness of the Higgs mass and of the cosmological constant as anthropic selection,
%and where predictability seems lost.}

This simple assumption  solves the two issues above and has strong consequences.

First, a quasi-flat inflationary potential is obtained because
the only adimensional potential is a quartic term $V(S) = \lambda_S |S|^4$ and
the only adimensional scalar/gravity coupling is $-\xi_S |S|^2 R$, so that $V_E = \bar M_{\rm Pl}^4 ( \lambda_S|S|^4)/(\xi_S|S|^2)^2 = \bar M_{\rm Pl}^4 \lambda_S/\xi_S^2$
is flat at tree level.
At quantum level the parameters $\lambda_S$ and $\xi_S$ run,
such that the slow-roll parameters are the beta-functions of the theory, as discussed in section~\ref{infl}.

Second, power divergences vanish just because of dimensional reasons:
they would have mass dimension, but there are no masses.
Vanishing of quadratic divergences leads to a modified version of naturalness, 
where the weak scale can be naturally small
even in absence of new physics at the weak scale~\cite{FN}
designed to protect the Higgs mass, such as supersymmetry or technicolor.

In this context
scale invariance is just an accidental symmetry, present at tree level because there are no masses.
Just like baryon number (a well known accidental symmetry of the Standard Model), scale invariance is
broken by quantum corrections.\footnote{Other attempts along similar lines assume
that scale or conformal invariance are exact symmetries at quantum level.  However, computable theories do not behave in this way.}
Then, the logarithmic
running of adimensional couplings can generate exponentially different scales via dimensional transmutation.
This is how the QCD scale arises.

The goal of this paper is exploring if the Planck scale and the electro-weak scale can arise in this context.

\subsubsection*{The theory}
The adimensional principle leads us to consider renormalizable
theories of quantum gravity
described by actions of the form:
\beq \label{eq:Ladim}
S = \int d^4x \sqrt{|\det g|} \,\bigg[ \frac{R^2}{6\gs^2} + \frac{\frac13 R^2 -  R_{\mu\nu}^2}{\gt^2} + 
\Lag^{\rm adim}_{\rm SM}+
\Lag^{\rm adim}_{\rm BSM}
\bigg].
\eeq
The first two terms, suppressed by the adimensional gravitational couplings $\gs$ and $\gt$,
are the graviton kinetic terms.\footnote{The second term is also known as `conformal gravity'.}
The third term,  $\Lag^{\rm adim}_{\rm SM}$, is
the adimensional part of the usual Standard Model (SM) Lagrangian:
\beq\label{eq:LadimSM}
 \Lag^{\rm adim}_{\rm SM} =  -\frac{F_{\mu\nu}^2}{4g^2}  + \bar \psi i \slashed{D} \psi + 
 |D_\mu H|^2 - (y H \psi\psi + \hbox{h.c.})  - \lambda_H  |H|^4 - \xi_H |H|^2 R 
\eeq
where $H$ is the Higgs doublet.
The last term,  $\Lag^{\rm adim}_{\rm BSM}$, describes possible new physics Beyond the SM (BSM).
For example adding a scalar singlet $S$ one would have
\beq\label{eq:LadimBSM}
 \Lag^{\rm adim}_{\rm BSM} = |D_\mu S|^2  - \lambda_S  |S|^4 + \lambda_{HS} |S|^2|H|^2 -\xi_S |S|^2 R.
\eeq
We ignore topological terms.
Non renormalizable terms,
the Higgs mass term $\frac12 M_h^2 |H|^2$
and the Einstein-Hilbert term $ -M_{\rm Pl}^2 R/16\pi$
are not present in the agravity Lagrangian, because they need dimensionful parameters.
The Planck mass can be generated dynamically if, at quantum level,
$S$ gets a vacuum expectation value such that 
$\xi_S \langle S\rangle^2 = M_{\rm Pl}^2/16\pi $~\cite{cosmon}.
The adimensional parameters of a generic agravity in 3+1 dimensions theory are:
\begin{enumerate}
\item the two gravitational couplings $\gs$ and $\gt$;
\item  quartic scalar couplings $\lambda$;
\item  scalar/scalar/graviton couplings $\xi$;
\item  gauge couplings $g$; 
\item  Yukawa couplings $y$.\footnote{The list would be much shorter
for $d\neq 4$.   Gauge couplings are adimensional only at $d=4$.
Adimensional scalar self-interactions exist at $d=\{3,4,6\}$.
Adimensional interactions between fermions and scalars exist at $d=\{3,4\}$. 
Adimensional fermion interactions exist at $d=2$.
%The $\xi$ term exists in any dimension. Other gravitational terms and scalar/gravitational terms exists in any even dimension.
}
\end{enumerate}
The graviton $g_{\mu\nu}$ has dimension zero, and eq.\eq{Ladim} is the most generic adimensional action compatible with general relativistic invariance.  The purely gravitational action just contains two terms:
the squared curvature $R^2$ and the Weyl term $\frac13 R^2 -  R_{\mu\nu}^2$.
They are suppressed by two constants, $\gs^2$ and $\gt^2$, that are the true adimensional gravitational couplings, in analogy to the gauge couplings $g$ that suppress the kinetic terms for vectors,
$-\frac14 F_{\mu\nu}^2/g^2$.
Thereby, the  gravitational kinetic terms contain 4 derivatives, and the 
graviton propagator is proportional to $1/p^4$.
Technically, this is how gravity becomes renormalizable.
In presence of an induced Planck mass, the graviton propagator becomes
\beq 
\frac{1}{M^2_2 p^2 - p^4} = \frac{1}{M^2_2} \bigg[\frac{1}{p^2} - \frac{1}{p^2 - M^2_2}\bigg]
\eeq
giving rise to a massless graviton with couplings suppressed by the Planck scale,
and to a spin-2 state with mass $M_2^2 = \frac12 \gt^2 \bar M_{\rm Pl}^2$ and negative norm.
Effectively, it behaves as an anti-gravity Pauli-Villars regulator for gravity~\cite{Stelle}. 
The Lagrangian can be rewritten
in a convoluted form where this is explicit~\cite{Ovrut} (any field with quartic derivatives can be 
rewritten in terms of two fields with two derivatives).  The $\gs$ coupling
gives rise to a spin-0 graviton with positive norm and mass $M_0^2 = \frac12 \gs^2 \bar M_{\rm Pl}^2 +\cdots$.
Experimental bounds are satisfied as long as $M_{0,2}\circa{>}\eV$.

At classical level,
theories with higher derivative suffer the Ostrogradski instability:
the Hamiltonian is not bounded from below~\cite{Ostro}.
At quantum level, creation of negative energy can be reinterpreted as destruction of positive energy:
the Hamiltonian becomes positive, but some states have negative norm and are called `ghosts'~\cite{Pais}.
This quantization choice amounts to adopt the same $i\epsilon$ prescription for the graviton and for the anti-graviton,
such that the cancellation that leads to renormalizability takes place.

We do not address the potential problem of a negative contribution to the cross-section for
producing an odd number of anti-gravitons with mass $M_2$ above their kinematical threshold.
Claims in the literature are controversial~\cite{Mannheim}.
Sometimes in physics we have the right equations
before having their right interpretation.
In such cases the strategy that pays is:
proceed with faith, %~\cite{fascistisumarte},
explore where the computations lead, %~\cite{Feynman},
if the direction is right the problems will disappear.%~\cite{Einstein}.

We  here compute the one loop quantum corrections of agravity, to
explore its quantum behaviour.
Can the Planck scale be dynamically generated?
Can the weak scale be dynamically generated?
%Can the theory be extrapolated up to infinite energy?

\section{Quantum  agravity}\label{RGE}

The quantum corrections to a renormalizable theory are mostly encoded in the renormalization group equations (RGE) for its parameters.
Ignoring gravity, a generic adimensional theory of real scalars $\phi_a$, Weyl fermions $\psi_j$ and vectors $V_A$
can be written as
\beq \Lag = - \frac14 (F_{\mu\nu}^A)^2 + \frac{(D_\mu \phi_a)^2}{2}  + \bar\psi_j i\slashed{D} \psi_j -
{\frac12}
(Y^a_{ij} \psi_i\psi_j \phi_a + \hbox{h.c.}) - \frac{\lambda_{abcd}}{4!} \phi_a\phi_b\phi_c\phi_d\label{eq:Lgen}\eeq
and its RGE have been computed up to 2 loops~\cite{MV}.
We here compute the one-loop $\beta$ functions $\beta_p \equiv dp/d\ln\mub$ of all parameters $p$
of a generic agravity theory, obtained adding to\eq{Lgen}
the generic scalar/graviton coupling \beq - \frac{\xi_{ab}}{2} \phi_a \phi_b R\eeq
as well as the graviton kinetic terms of eq.\eq{Ladim} and the minimal gravitational interactions
demanded by general relativity.

Previous partial computations found contradictory results and have been performed with ad hoc techniques in a generic background.  We instead follow the usual Feynman diagrammatic approach,\footnote{One of the authors (A. Salvio) adapted the  public tools~\cite{Feynart}; the other author (A. Strumia) employed his own equivalent codes.} expanding around a flat background (the background is just an infra-red property which does not affect ultra-violet divergences).

\subsection{The graviton propagator}
Eq.\eq{Ladim} is the most general action containing adimensional powers of the fundamental fields.
Concerning the purely gravitational sector, apparently there are extra terms such as $D^2 R$ or $R_{\mu\nu\alpha\beta}^2$.
However the first one is a pure derivative; and the second one can be eliminated using the
topological identity
\beq R_{\alpha\beta\mu\nu}^2 - 4 R_{\mu\nu}^2 + R^2 = \frac{1}{4}\epsilon^{\mu\nu\rho\sigma}
\epsilon_{\alpha\beta\gamma\delta}R_{\mu\nu}^{\alpha\beta} R_{\rho\sigma}^{\gamma\delta}\simeq 0.
 \eeq
The combination suppressed by $\gt^2$ in eq.\eq{Ladim} is the square of the Weyl or conformal tensor,
defined by subtracting all traces to the Riemann tensor:
\beq W_{\mu\nu\alpha\beta} \equiv R_{\mu\nu\alpha\beta} + \frac{1}{2} (
g_{\mu\beta} R_{\nu\alpha} -g_{\mu\alpha} R_{\nu\beta}+ g_{\nu\alpha} R_{\mu\beta} - g_{\nu\beta}R_{\mu\alpha})
+\frac16 (g_{\mu\alpha}g_{\nu\beta}-g_{\nu\alpha}g_{\mu\beta} )R.
\eeq
Indeed
\beq  \frac12 W_{\alpha\beta\mu\nu}^2  =  \frac12 R_{\alpha\beta\mu\nu}^2 - R_{\mu\nu}^2 + \frac16 R^2 \simeq
R_{\mu\nu}^2 - \frac13 R^2. \eeq
We expand around the flat-space metric $\eta_{\mu\nu}=\diag(1,-1,-1,-1)$ as $g_{\mu\nu} = \eta_{\mu\nu} + h_{\mu\nu}$ such that
\beq R\sqrt{|\det g|} =  (\partial_\mu\partial_\nu - \eta_{\mu\nu} \partial^2)h_{\mu\nu} + \frac14 (h_{\mu\nu}\partial^2 h_{\mu\nu}-h_{\alpha\alpha}\partial^2 h_{\alpha\alpha}+2 h_{\mu\nu}\partial_\mu\partial_\nu  h_{\alpha\alpha}-2 h_{\mu\alpha}
\partial_\alpha\partial_\beta h_{\beta\mu})+\cdots.\eeq
%\footnote{We use notations similar to~\cite{Stelle}, but we use mostly negative metric, perform a different expansion, and choose a simpler gauge,
%such that our action is written in terms of projectors, see eq.\eq{Sg2}.}
To quantise the theory we follow the Fadeev-Popov procedure adding the gauge fixing term
\beq \label{eq:gf}
S_{\rm gf} = - \frac{1}{2\xi_g}\int d^4x ~ f_\mu \partial^2 f_\mu,\qquad
f_\mu = \partial_\nu h_{\mu\nu}.
\eeq
We choose a non-covariant term quadratic in $h_{\mu\nu}$, such that gauge fixing does not affect the graviton couplings.
%For $\xi_g=1$ it gives an analogous of the de Donder gauge, adapted for a quartic propagator.
At quadratic level the purely gravitational action is
\begin{eqnarray} S &=& \frac12 \int d^4k~k^4~ {h}_{\mu\nu}  \bigg[ -\frac{1}{2\gt^2} P^{(2)}_{\mu\nu\rho\sigma} 
+ \frac{1}{\gs^2} P^{(0)}_{\mu\nu\rho\sigma}  + \frac{1}{2\xi_g} (P^{(1)}_{\mu\nu\rho\sigma}+2{P}^{(0w)}_{\mu\nu\rho\sigma})
 \bigg]  h_{\rho\sigma}\label{eq:Sg2}
\end{eqnarray}
where 
\begin{eqnsystem}{sys:P}
P^{(2)}_{\mu\nu\rho\sigma} &=& \frac12 T_{\mu\rho}T_{\nu\sigma} + \frac12 T_{\mu\sigma} T_{\nu\rho}-\frac{T_{\mu\nu}T_{\rho \sigma}}{d-1} \\
P^{(1)}_{\mu\nu\rho\sigma} &=& \frac12 (T_{\mu\rho}L_{\nu\sigma} +  T_{\mu\sigma} L_{\nu\rho}+T_{\nu\rho}L_{\mu\sigma} +  T_{\nu\sigma} L_{\mu\rho})\\
P^{(0)}_{\mu\nu\rho\sigma} &=& \frac{T_{\mu\nu}T_{\rho \sigma}}{d-1} \\
 P^{(0w)}_{\mu\nu\rho\sigma} &=& L_{\mu\nu}L_{\rho \sigma}
 \end{eqnsystem}
are projectors over spin-2, spin-1 and spin-0 components of $h_{\mu\nu}$, written in terms of $d=4-2\epsilon$,
$T_{\mu\nu} = \eta_{\mu\nu} - k_\mu k_\nu/k^2$ and $ L_{\mu\nu}=k_\mu k_\nu/k^2$.
Their sum equals unity: $(P^{(2)}+P^{(1)}+P^{(0)}+P^{(0w)})_{\mu\nu\rho\sigma}=\frac12 (\eta_{\mu\nu}\eta_{\rho\sigma}+\eta_{\mu\sigma}\eta_{\rho\nu})$.
Inverting the kinetic term of eq.\eq{Sg2} we find the graviton propagator
\beq 
D_{\mu\nu\,\rho\sigma} = \frac{i}{k^4}
 \bigg[ -2\gt^2  P^{(2)}_{\mu\nu\rho\sigma}  + \gs^2 P^{(0)}_{\mu\nu\rho\sigma} + 2\xi_g (P^{(1)}_{\mu\nu\rho\sigma}  + \frac12 P^{(0w)}_{\mu\nu\rho\sigma}  )
  \bigg] .
\eeq

\subsubsection*{Gravitational ghost couplings}
One needs to path-integrate over Fadeev-Popov ghosts $\eta_\alpha$ and $\bar\eta_\mu$ with action
\beq 
S_{\rm ghost}= \int d^4x \,d^4y\, \bar{\eta}_\mu(x) \frac{\delta f_\mu(x)}{\delta \xi_\alpha(y)} \eta_\alpha(y) .\eeq
By performing an infinitesimal
transformation $x_\mu \to x'_\mu=x_\mu+\xi_\mu(x)$ one finds the transformation of $h_{\mu\nu}$
at first order in $\xi_\mu$ and at a fixed point $x_\mu$:
\beq \delta h_{\mu\nu}= - (\partial_\mu \xi_\nu + \partial_\nu \xi_\mu) - (h_{\alpha\mu}\partial_\nu +h_{\alpha\nu}\partial_\mu + (\partial_\alpha h_{\mu\nu}))\xi_\alpha.\eeq
The ghost action then is
\beq
S_{\rm ghost} = 
\int d^4x\, \{\partial_\alpha\bar{\eta}_\mu (\partial_\alpha  \eta_\mu +\partial_\mu \eta_\alpha) +\partial_\nu\bar{\eta}_\mu[h_{\alpha\mu}\partial_\nu \eta_\alpha+ h_{\alpha\nu}\partial_\mu \eta_\alpha+
 (\partial_\alpha h_{\mu\nu}) \eta_\alpha ] \}.
\eeq
In order to verify gravitational gauge-independence we will perform all computations using a more general gauge fixing, given by eq.\eq{gf} with  $f_\mu = \partial_\nu( h_{\mu\nu} - c_g \frac12 \eta_{\mu\nu} h_{\alpha \alpha})$.\footnote{The graviton propagator becomes
$$
D_{\mu\nu\,\rho\sigma} =
\frac{i}{k^4} \bigg[ -2 \gt^2 P^{(2)} +\gs^2 \bigg(P^{(0)}+\frac{\sqrt{3}  c_g T^{(0)}}{2-c_g} 
+\frac{3c_g^2P^{(0w)} }{(2-c_g)^2} \bigg)
+2\xi_g \bigg( P^{(1)}+\frac{2 P^{(0w)}}{(2-c_g)^2}\bigg)
  \bigg]_{\mu\nu\rho\sigma} 
$$
where $T^{(0)}  _{\mu\nu\rho\sigma} =(T_{\mu\nu}L_{\rho\sigma} + L_{\mu\nu}T_{\rho\sigma})/\sqrt{d-1}$.  
The gauge of eq.\eq{gf} corresponds to $c_g=0$; the gauge used in~\cite{Stelle} corresponds to $c_g=1$, which is a convenient choice in Einstein gravity.  
In the  generic gauge $c_g\neq0$  the ghost Lagrangian is
$$
\bar{\eta}_\mu \bigg [\partial^2 \eta_{\alpha\mu} + (1-c_g)\partial_\alpha \partial_\mu+
\partial_\nu h_{\alpha\mu}\partial_\nu+\partial_\nu h_{\alpha\nu}\partial_\mu+
\partial_\nu (\partial_\alpha h_{\mu\nu})- c_g\partial_\mu h_{\alpha\nu}\partial_\nu-\frac{c_g}{2}\partial_\mu (\partial_\alpha h_{\nu\nu}) \bigg] \eta_\alpha
$$
so that the ghost propagator is 
$$-\frac{i}{k^2}\bigg[ \eta_{\mu\nu}+ \frac{c_g-1}{2-c_g} \frac{k_\mu k_\nu}{k^2}\bigg].$$}

\smallskip

We fix  gauge invariance of vectors $V_\mu$ adding to the Lagrangian the standard $\xi$-gauge
term $-f_V^2/2(1-\xi_V)$ with $f_V = \partial_\mu V_\mu$, such that the vector propagator
is $i(-g_{\mu\nu}+\xi_V k_\mu k_\nu/k^2)/k^2$.
Such term does not depend on gravitons, so that the gauge-invariances of spin-2 and spin-1 particles
are fixed independently.

\subsection{Wave-function renormalizations}\label{wave}
We write the gauge-covariant derivatives as $D_\mu \phi_a= \partial_\mu \phi_a+ i \theta^A_{ab} V^A_\mu \phi_b$ when acting on scalars and as $D_\mu\psi_j = \partial_\mu \psi_j + i t^A_{jk}V^A_\mu\psi_k$ when acting on fermions
(the gauge couplings are contained in the matrices $\theta^A$ and $t^A$).
In this paper we adopt dimensional regularization and the modified minimal subtraction renormalization scheme
$\overline{\rm MS}$, with energy scale $\bar\mu$.
The anomalous dimensions $\gamma$ of the fields are defined as $\gamma = \frac12 d\ln Z/d\ln\mub$ in terms of
the wave-function renormalization constants $Z = 1 + \delta Z$.
Gravitational couplings of fermions are derived following the formalism of~\cite{Woodard}.
By computing the one-loop corrections to the matter kinetic terms
we find the one-loop anomalous dimension of scalars
\bea\label{eq:ZS}
(4\pi)^2 \gamma^S_{ab} &=& \Tr\, Y^a Y^{\dagger b} - (2+\xi_V) \theta^A_{ac} \theta^A_{cb} + \\
&&+\gs^2 \left(\frac{3 \left(c_g-1\right){}^2}{4 \left(c_g-2\right){}^2}  \delta_{ab}  +\frac{3c_g \xi_{ab}  }{c_g-2}\right)
+\frac{3 c_g^2-12
   c_g+13}{4 (c_g-2)^2}  \xi _g
\delta_{ab}
\nonumber  \\
\riga{and of fermions}\\
(4\pi)^2 \gamma^F &=&\frac12 Y^a Y^{\dagger a} + (1-\xi_V) t^A t^A+  \label{eq:Zf}\\
&& -\frac{25}{16}\gt^2+\gs^2 \frac{22-16c_g+7 c_g^2}{16(c_g-2)^2} + 3\xi_g\frac{22-20c_g+5 c_g^2}{16(c_g-2)^2}\nonumber
\eea
where the first lines show the well known matter terms~\cite{MV},
and the second lines show the new terms due to agravity, as computed for generic gauge-fixing parameters $\xi_g$ and $c_g$.
Vectors are discussed in the next section.

\begin{figure}
\begin{center}
\includegraphics[width=0.65\textwidth]{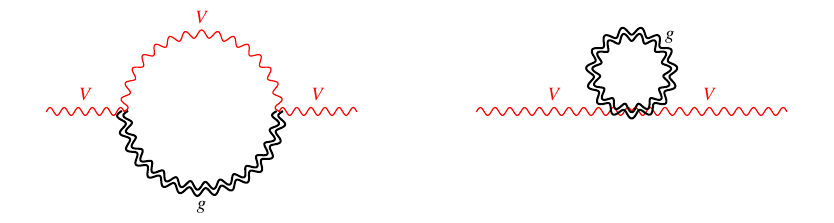}
\caption{\label{fig:VV}\em Gravitational corrections to the running of the gauge couplings.}
\end{center}
\end{figure}

\subsection{RGE for the gauge couplings}
The one-loop correction to the kinetic term of vectors describes the RGE for the gauge couplings.
The two new gravitational Feynman diagrams are shown in figure\fig{VV}.
They have opposite logarithmic divergences, so that their sum is finite.
This cancellation was noticed in~\cite{Frad,Narain} in more general contexts, and seems due to the fact that gravitons
have no gauge charge.\footnote{Other authors try to interpret ambiguous power-divergent
corrections to gauge couplings from Einstein gravity as gravitational power-running RGE, 
with possible physical consequences such as an asymptotically free hypercharge~\cite{SW}.
We instead compute the usual unambiguous logarithmic running, in the context of theories where power divergences vanish.}

In conclusion, the one-loop RGE for the gauge couplings do not receive any gravitational correction and the 
usual one-loop RGE for the gauge couplings remain valid also above the Planck scale.
Within the SM, the hypercharge gauge couplings hits a Landau pole at $\bar\mu\sim 10^{41}\GeV$.

\subsection{RGE for the Yukawa couplings}
\begin{figure}[t]
\begin{center}
\includegraphics[width=\textwidth]{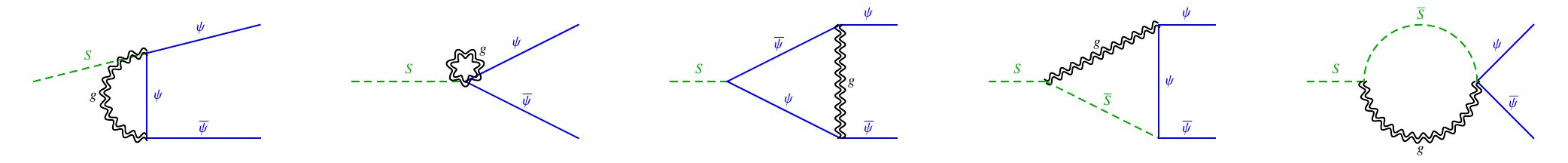}
\caption{\label{fig:ffS}\em Gravitational corrections to the running of the Yukawa couplings.}
\end{center}
\end{figure}

Summing the diagrams of fig.\fig{ffS} gives a divergent correction
%\beq \frac{Y^a}{(4\pi)^2 \epsilon}\bigg[-\frac52 \gt^2 +\gs^2 \frac{28+c_g^2(12-24\xi_S)+4 c_g(12\xi_S-7)}{16(c_g-2)^2}+
%\xi_g \frac{92-84c_g+21 c_g^2}{12(2-c_g)^2} \bigg]\eeq
that depends on the gauge fixing parameters and on the scalar/graviton couplings $\xi$.
Adding the fermion and scalar wave function renormalizations of section~\ref{wave}
such dependencies cancel.
We find the one-loop RGE:
\bea\label{eq:RGEY}
(4\pi)^2 \frac{dY^a}{d\ln\mub} &=& \frac12(Y^{\dagger b}Y^b Y^a + Y^a Y^{\dagger b}Y^b)+ 2 Y^b Y^{\dagger a} Y^b + \nonumber\\
&&+ Y^b \Tr(Y^{\dagger b} Y^a) - 3 \{ C_{2F} , Y^a\}  + \frac{15}{8}\gt^2 Y^a.~~~
\eea
where $C_{2F} = t^A t^A$, and the latter term is the contribution due to agravity and has the opposite sign with respect to the analogous
multiplicative term due to gauge interactions.
Specializing eq.\eq{RGEY} to the SM, we find the one-loop RGE for the top quark Yukawa coupling:
\bea
(4\pi)^2 \frac{dy_t}{d\ln\mub} &=& \frac{9}{2} y_t^3 + y_t (\frac{15}{8} \gt^2 - 8 g_3^2 - \frac94g_2^2-\frac{17}{20} g_1^2).
\eea
We know of no previous computation in the literature.

\subsection{RGE for the quartic couplings}
Tens of Feynman diagrams contribute to the scalar 4-point function at one loop. After summing them and taking into account the scalar
wave-function renormalization of eq.\eq{ZS} the gauge dependence disappears and we find the one-loop RGE:
\bea
(4\pi)^2 \frac{d\lambda_{abcd}}{d\ln\mub} &=&  \sum_{\rm perms} \bigg[\frac18
  \lambda_{abef}\lambda_{efcd}+
  \frac38 \{\theta^A,\theta^B\}_{ab}\{\theta^A ,\theta^B\}_{cd}
  -\Tr\, Y^a Y^{\dagger b} Y^c Y^{\dagger d}+
   \nonumber
\\
% &&+
%3 [\{\theta^A,\theta^B\}_{ab}\{\theta^A ,\theta^B\}_{cd}+
%\{\theta^A,\theta^B\}_{ac}\{\theta^A ,\theta^B\}_{bd}+
%\{\theta^A,\theta^B\}_{ad}\{\theta^A ,\theta^B\}_{bc}]+  \nonumber \\
&&\label{eq:RGElambda}
+\frac58 \gt^4 \xi_{ab}\xi_{cd}+  \frac{ \gs^4}{8}\xi_{ae}\xi_{cf}(\delta_{eb}+6\xi_{eb})(\delta_{fd}+6\xi_{fd}) +\\
&&+\frac{f_0^2}{4!}  (\delta_{ae}+6\xi_{ae})(\delta_{bf}+6\xi_{bf})\lambda_{efcd}\bigg]
+ \lambda_{abcd} \bigg[ \sum_k (Y_2^k-3  C_{2S}^k)+ 5 \gt^2\bigg]   ,\nonumber
%+\frac38 \sum_{\rm perms} \{ \theta^A,\theta^B\}_{ab} \{ \theta^A,\theta^B\}_{cd}
\eea
where the first sum runs over the $4!$ permutations of $abcd$ and
the second sum over $k=\{a,b,c,d\}$, with $Y_2^k$ and $C_{2S}^k$ defined  by 
\be {\rm Tr} (Y^{\dagger a}Y^b) =Y_2^a \delta ^{ab}, \quad \theta^A_{ac} \theta^A_{cb}= C_{2S}^{a} \delta_{ab}.\ee 
RGE for quartics have been computed in the literature in some models~\cite{EOR}; 
we find a simpler result where the $\gt^2$ term does not depend on $\xi$ and where the
$\gs^2$ term vanishes if $\xi_{ab} = -\delta_{ab}/6$.

Specializing our general  eq.\eq{RGElambda} to the case of the SM
Higgs doublet plus a complex scalar singlet $S$ with action given by eq.\eq{Ladim}, the RGE become:
\bea
(4\pi)^2 \frac{d\lambda_S}{d\ln\mub} &=& 20\lambda_S^2 + 2\lambda_{HS}^2 + \frac{\xi_S^2}{2} \left(5 \gt^4+\gs^4 (1+6\xi_S)^2\right) + \lambda_S \left(5 \gt^2+\gs^2(1+6\xi_S)^2\right),~\\
(4\pi)^2 \frac{d{\lambda_{HS}}}{d\ln\mub} &=& -\xi_H\xi_S\left( 5\gt^4  + \gs^4(6\xi_S+1)(6\xi_H+1)\right) - 4\lambda_{HS}^2+\lambda_{HS}\bigg(8\lambda_S + 12 \lambda_H + 6 y_t^2+  \nonumber \\
\label{eq:RGElambdaHS}
&& +5 \gt^2+\frac{\gs^2}{6}\left[(6\xi_S+1)^2+(6\xi_H+1)^2+4(6\xi_S+1)(6\xi_H+1)\right]  \bigg),   \\
(4\pi)^2 \frac{d\lambda_H}{d\ln\mub} &=& \frac{9}{8}g_2^4+\frac{9}{20}g_1^2g_2^2+\frac{27}{200}g_1^4-6 y_t^4+24\lambda_H^2+\lambda_{HS}^2+\frac{\xi_H^2}{2}\left(5 \gt^4+\gs^4(1+6\xi_H)^2\right)+\nonumber
\\
&&+\label{eq:RGElambdaH}
\lambda_{H}\left(5 \gt^2+\gs^2 (1+6\xi_H)^2+12 y_t^2- 9 g_2^2 - \frac95 g_1^2 \right).
\eea

\subsection{RGE for the scalar/graviton couplings}\label{RGExi}
We extract the one-loop  RGE for the $\xi$ parameters from the one-loop correction 
to the 
 \\ graviton$_{\mu\nu}$/scalar/scalar vertex.  At tree level, two different Lagrangian terms contribute to such vertex: 
\begin{itemize}
\item[a)] one contribution comes from the scalar kinetic term
(when the graviton momentum is vanishing and the two scalars have momenta $\pm p$);
\item[b)] one contribution comes from $\xi$ terms (when the graviton has a non-vanishing momentum $k$ and one scalar has zero momentum).
\end{itemize}
We compute both contributions.
We find that the correction a) reproduces the scalar wave-function renormalization already computed in section~\ref{wave},
including the correct tensorial structure $\frac12 p^2 \eta_{\mu\nu} - p_\mu p_\nu$, provided that the 
graviton field is renormalized as follows:
\beq    h_{\mu\nu} \to \frac{1}{\sqrt{Z_{TL}}} \bigg(h_{\mu\nu}-\frac{1}{4}\eta_{\mu\nu} h_{\alpha\alpha}\bigg)+
\frac{1}{\sqrt{Z_{T}}} \frac{1}{4}\eta_{\mu\nu} h_{\alpha\alpha}\label{eq:Zg}\eeq
%\beq h_{\alpha\alpha}  \rightarrow \frac{h_{\alpha\alpha}}{\sqrt{Z_{T}}}, \quad h_{\mu\nu}^{TL} \rightarrow \frac{h_{\mu\nu}^{TL}}{\sqrt{Z_{TL}}}, \label{eq:Zg}\eeq
%where $h_{\mu\nu}^{TL}$ is the traceless part of the graviton field 
%%
%\be h_{\mu\nu}^{TL}\equiv h_{\mu\nu}-\frac{1}{4}\eta_{\mu\nu} h_{\alpha\alpha}.\ee
%%
The wave-function renormalization $Z_T$ and $Z_{TL}$ differ 
because we used a simple gravitational gauge-fixing term that breaks general relativity but respects special
relativity: thereby distinct representations of the Lorentz group
(the trace and the traceless part of $h_{\mu\nu}$), 
get different renormalizations.
We find the one-loop results
\bea
Z_T &=& 1 + \frac{1}{(4\pi)^2\epsilon}  \frac{12 c_g-13-3c_g^2}{(c_g-2)^2}\xi_g,\\
Z_{TL}&=& 1 + \frac{1}{(4\pi)^2\epsilon}\bigg[\frac{10}{9}\frac{c_g-4}{c_g-2} \gt^2-\frac29 \frac{4-3c_g+2c_g^2}{(c_g-2)^2}\gs^2 -\frac23 \frac{9-8c_g+2c_g^2}{(c_g-2)^2}\xi_g
\bigg].~~~~~
\eea
%Defining $  Z_{T} \equiv 1+\delta Z_T$ and $  Z_{TL}\equiv 1+\delta Z_{TL} $ we find
%%
%\be (4\pi)^2 \delta Z_{T} =   \{  \frac{4\xi_g}{  (d/2-2)} ,\frac{-13\xi_g}{ 4  (d/2-2)}  \}, \quad  (4\pi)^2 \delta Z_{TL} =   \{  \frac{2(f_0^2-5f_2^2+3\xi_g)}{3 (d/2-2)} , \frac{2f_0^2-20f_2^2-27\xi_g/2}{9 (d/2-2)} \}\ee
%%
%for $c_g=1$ and $c_g=0$ respectively.
We verified that we find the same $Z_{TL}$ by computing the graviton renormalization constant from
the one-loop graviton/vector/vector vertex.

\medskip

Next, we compute the correction b).  After adding to it the scalar and the graviton wave-function renormalization,
we find that the total correction has the correct tensorial structure $k^2 \eta_{\mu\nu} - k_\mu k_\nu$ and corresponds to the following one-loop RGE for the $\xi$ parameters:
\bea
(4\pi)^2 \frac{d\xi_{ab}}{d\ln\mub} &=& \frac16\lambda_{abcd}\left(6\xi_{cd}+\delta_{cd}\right) 
+ (6\xi_{ab}+\delta_{ab})   \sum_k \left[\frac{Y_2^k}{6} - \frac{C_{2S}^k}{2}  \right]+\nonumber \\
&& -\frac{5f_2^4}{3f_0^2} \xi_{ab}
+f_0^2  \xi_{ac}\left( \xi_{cd}+\frac23 \delta_{cd}\right)(6 \xi_{db}+\delta_{db})\label{eq:RGExi}
\eea
where the sum runs  over $k=\{a,b\}$.
RGE for the $\xi$ term have been computed in the literature in some  models~\cite{Shapiro,Yoon2,EOR}; we find different and simpler gravitational terms.
%The unusual factor $f_0^2$ at the denominator would disappear if, instead of writing
%the RGE for the conventional $\xi_{ab}$ couplings, we would instead employ the natural parameter
%$\xi'_{ab} =\gs^2 \xi_{ab}$.  NO RIAPPARE DA xi = xi'/f0^2

Specialising our general  eq.\eq{RGExi} to the case of the SM
Higgs doublet plus a complex scalar singlet $S$ with action given by eq.\eq{Ladim}, the RGE become:
\bea
(4\pi)^2 \frac{d\xi_S}{d\ln\mub} &=& (1+6\xi_S) \frac{4}{3}\lambda_S
-\frac{2\lambda_{HS}}{3} (1+6\xi_H)
+\frac{\gs^2}{3}\xi_S(1+6\xi_S)(2+3\xi_S) - \frac53 \frac{\gt^4}{\gs^2}\xi_S,\\
(4\pi)^2 \frac{d\xi_H}{d\ln\mub} &=& (1+6\xi_H)(y_t^2-\frac34 g_2^2 - \frac{3}{20} g_1^2+2\lambda_H)
-\frac{\lambda_{HS}}{3} (1+6\xi_S)+  \nonumber \\
&&+\frac{\gs^2}{3}\xi_H(1+6\xi_H)(2+3\xi_H) - \frac53 \frac{\gt^4}{\gs^2}\xi_H.
\label{eq:RGExiHS}
\eea

\subsection{RGE for the agravitational couplings}
The RGE for the  couplings $\gs,\gt$ are computed summing
the one loop corrections to the graviton kinetic term at 4th order in the external momentum $k$ from:
a) the graviton rainbow and seagull diagrams; 
b) the gravitational ghost;
c) the graviton wave function renormalization of eq.\eq{Zg}.
After combining all these ingredients we find a tensorial structure equal to the structure of the graviton kinetic term of eq.\eq{Sg2}
that thereby can be interpreted as a renormalization of $\gs$ and $\gt$.
Adding also the matter contributions (that separately have the correct tensorial structure) we obtain:
\begin{eqnsystem}{sys:RGG}
(4\pi)^2\frac{d \gt^2}{d\ln\mub}&=& -\gt^4\bigg(\frac{133}{10} +\frac{N_V}{5}+\frac{N_f}{20}+\frac{N_s}{60}
\bigg),\\
(4\pi)^2\frac{d \gs^2 }{d\ln\mub}&=&  \frac53 \gt^4 + 5 \gt^2 \gs^2 + \frac56 \gs^4 +\frac{\gs^4}{12} (\delta_{ab}+6\xi_{ab})(\delta_{ab}+6\xi_{ab}).
\end{eqnsystem}
Here $N_V$, $N_f$, $N_s$ are the number of vectors, Weyl fermions and real scalars.
In the SM $N_V=12$, $N_f = 45$, $N_s = 4$.
Unlike in the gauge case, all the contributions to the RGE of $\gt$ have the same sign, such that $\gt$
is always asymptotically free.
This result agrees with~\cite{Frad,Avramidi}.
The RGE for $f_0$ agrees with~\cite{Avramidi}.
See~\cite{Shapiro,Narain} for results with different signs.
The matter contributions had been computed in~\cite{gravmat} leading to the concept of `induced gravity'~\cite{InducedGravity},
which in our language corresponds to the RGE running of $\gt$.
Concerning the pure gravitational effect, agravity differs from gravity.
The coupling $\gs$ is asymptotically free only for $\gs^2<0$ which leads to a tachionic instability $M_0^2<0$.

\section{Dynamical generation of the Planck scale}\label{PlDyn}
Having determined the quantum behaviour of agravity, we can now study if the Planck scale can be generated dynamically.
The possible ways are:
\begin{enumerate}
\item[a)] {\bf Non-perturbative}:
the couplings $\gt$ or $\xi$ become non-perturbative when running down to low energy:
the Planck scale can be generated in a way similar
to how the QCD scale is generated.

\item[b)] {\bf Perturbative}:  
a  quartic $\lambda_S$ runs in such a way that
$S$ gets a vacuum expectation value 
$\xi \langle S\rangle^2 = \bar M_{\rm Pl}^2/2 $.

\end{enumerate}
We focus on the second, perturbative, mechanism.
In the usual Coleman-Weinberg case, $S$ acquires a vev if its quartic $\lambda_S$ becomes negative
when running down to low energy, and $\langle S\rangle$ is roughly given by the RGE scale at which $\lambda(\bar\mu)$
becomes negative.
This can be understood by noticing that the quantum effective potential is roughly given by $V_{\rm eff} = \lambda_S(\bar\mu \approx S) |S|^4$.
%: it keep the tree-level form, but with the quartic coupling renormalized at a scale close to the vev.
This means that the vacuum energy is always negative, in contrast to the observed near-vanishing vacuum energy.

In the gravitational case the situation is different, precisely because the effective Lagrangian contains
the $-\frac12 f(S)  R$ term, that should generate the Planck scale.
The field equation for the scalar $S$ in the homogeneous limit is
\beq V' + \frac{f'(S)}{2}  R=0\eeq
and the (trace of the) gravitational equation is
\beq f R + 4V = {\cal O}(R^2/f_{0,2}^2)\eeq 
where, around the phenomenologically desired flat-space solution, we can neglect the $R^2$ term
with respect to the induced Einstein term.
By eliminating $R$ we obtain the minimum equation for $S$:
\beq V' - \frac{2f'}{f} V=0\eeq
or, equivalently,
\beq V'_E = 0 \qquad \hbox{where}\qquad V_E  =\bar M_{\rm Pl}^4  \frac{V}{f^2}
\label{eq:VE}\eeq
is called Einstein-frame potential because eq.\eq{VE} can be obtained by performing a 
field redefinition $g^E_{\mu\nu}= g_{\mu\nu}\times  f/\bar M_{\rm Pl}^2 $ such that the coefficient of $R_E$ in the Lagrangian has the canonical Einstein value.  Under this transformation\footnote{Quantum corrections change when changing frame.  In a generic context this leads to ambiguities.
In the agravity context the fundamental action is given by eq.\eq{Ladim}, thereby quantum corrections must be computed in the `Jordan frame'.}
 the Lagrangian for the modulus of the scalar $|S| = s /\sqrt{2}$  becomes
\beq \sqrt{\det g} \bigg[\frac{(\partial_\mu s)^2}{2} -  \frac{f}{2} R - V\bigg]  = \sqrt{\det g_E} \bigg[  K
\frac{(\partial_\mu s)^2}{2} - \frac{\bar M_{\rm Pl}^2}{2} R_E -V_E\bigg].\label{eq:LS}\eeq
% here I got the factors from 1402.2129  Giudice Lee
where
\beq  K = \bar M_{\rm Pl}^2\bigg(\frac{1}{f} +  \frac{3 f^{\prime 2}}{2 f^2}\bigg).\eeq
The non-canonical  factor $K$ in the kinetic term for $s$
can be reabsorbed by defining a canonically normalised Einstein-frame scalar $s_E(s)$ as
$ {ds_E}/{ds} = \sqrt{K}$
such that the Lagrangian becomes
\beq \label{eq:LsE}
\sqrt{\det g_E} \bigg[ \frac{(\partial_\mu s_E)^2}{2} - \frac{\bar M_{\rm Pl}^2}{2} R_E -V_E(s_E)\bigg].\eeq
In the agravity scenario $f(S)$ is approximatively given by
$f(S) = \xi_S(\bar \mu\approx s)s^2$ and $V_S$ by $\lambda_S(\bar\mu \approx s) s^4/4$.
Thereby the Einstein-frame potential is given by
\beq V_E(S) =\frac{\bar M_{\rm Pl}^4}{4} \frac{\lambda_S(s)}{\xi_S^2(s)}\eeq
and the vacuum equation is
\beq  \frac{\beta_{\lambda_S}(s)}{\lambda_S(s)} - 2 \frac{\beta_{\xi_S}(s)}{\xi_S(s)}=0\label{eq:agmin}\eeq
where $\beta_p = dp/d\ln_\mu$ are the $\beta$ functions of the couplings $p$.
This equation is significantly different from the analogous equation of the usual non-gravitational
Coleman-Weinberg mechanism, which is $\lambda_S(s)=0$.

\bigskip

Furthermore, we  want a nearly-vanishing cosmological constant.  

Unlike in the non-gravitational Coleman-Weinberg case, where $V$ is always negative at the minimum, in the  agravity context $V(s) = \lambda_S(s) s^4/4$ can be vanishing at the minimum, provided that $\lambda_S(s)=0$ at the minimum.
This equation has the same form as the Coleman-Weinberg minimum condition, but its origin is different:
it corresponds to demanding a negligible cosmological constant.

In summary, agravity can generate the Planck scale while keeping the vacuum energy vanishing provided that
\beq\left\{
\begin{array}{rcll}
\lambda_S(s) &=& 0 & \hbox{(vanishing cosmological constant),}\\
\beta_{\lambda_S}(s) &=& 0 & \hbox{(minimum condition),}\\
\xi_S(s) s^2 &=& \bar M_{\rm Pl}^2 & \hbox{(observed Planck mass).}
\end{array}\right.
\label{eq:agravMPl}
\eeq
The minimum equation of eq.\eq{agmin} has been simplified taking into account that $\lambda_S$ nearly vanishes at the minimum.
%\footnote{We are not considering the possibility that instead $\xi_S$ runs to large non-perturbative values.
%\xxx{suppressed cosm cte?  Discuss here $\xi_s\sim 10^{30}$}
%}

In the present scenario the cosmological constant can be naturally suppressed down to about $M_0^2$:
even making it as light as possible, $M_0\sim \eV$, the cosmological constant is
at least $60$ orders of magnitude larger than the observed value.
Thereby we just invoke a huge fine-tuning without trying to explain the smallness of the cosmological constant.

\begin{figure}[t]
\begin{center}
\includegraphics[width=0.95\textwidth]{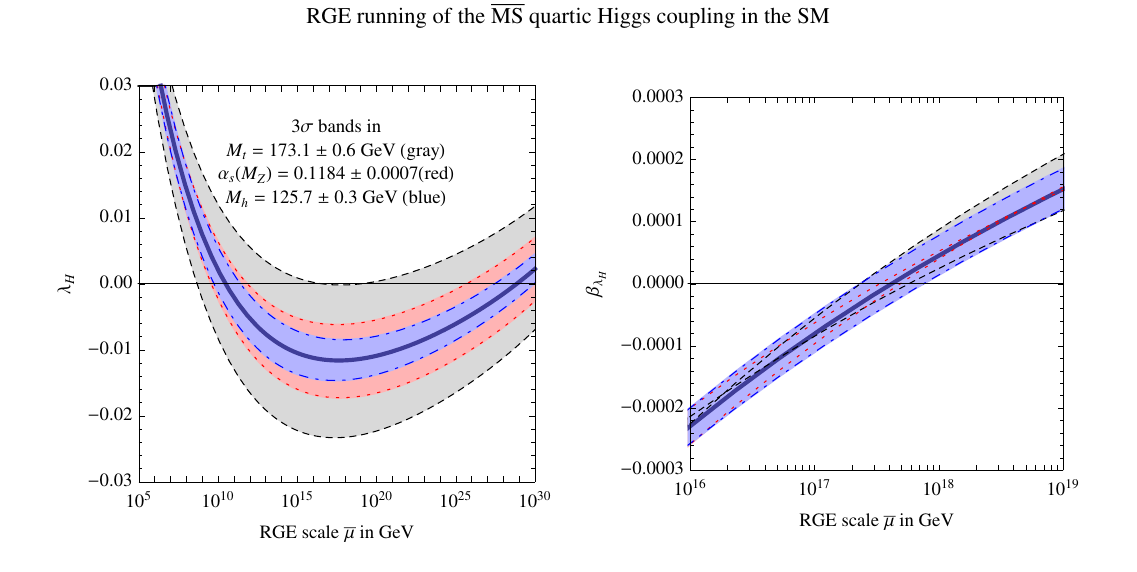}
\caption{\label{fig:RGESM}\em Running of the quartic Higgs coupling in the SM~\cite{NNLO}.
Agravity corrections can increase $\beta_{\lambda_H} = d\lambda_H/d\ln\mub$ and thereby $\lambda_H$ at scales above $M_{0,2}$.
Parameterizing the effective potential as $V_{\rm eff}(h) \equiv \lambda_{\rm eff}(h) h^4/4$,
the $\beta$ function of $\lambda_{\rm eff}$ vanishes at a scale a factor of few higher than 
the $\beta$ function of $\lambda_H$.}
\end{center}
\end{figure}

\subsection*{Models}\label{mod}
In words, the quartic $\lambda_S$ must run in such a way that it vanishes together with its $\beta$ function around the Planck mass.  Is such a behaviour possible?
The answer is yes; for example this is how the Higgs quartic $\lambda_H$ can run in the Standard Model (see fig.\fig{RGESM}a, upper curve).
Its $\beta$ function vanishes at the scale where the gauge coupling contribution to $\beta_{\lambda_H}$ in eq.\eq{RGElambdaH} compensates the top Yukawa contribution.
Fig.\fig{RGESM}b shows that this scale happens to be close to the Planck mass.
Although we cannot identify the Higgs field with the $S$ field --- the Higgs vev is at the weak scale
so that the possible second minimum of the Higgs potential at the Planck mass is not realised in nature ---
the fact that the conditions of eq.\eq{agravMPl} are realised in the SM is encouraging
in showing that they can be realised and maybe points to a deeper connection.

By considering the generic RGE of agravity, one can see that in the pure gravitational limit 
the conditions of eq.\eq{agravMPl} cannot be satisfied, so the scalar $S$ must have extra gauge
and Yukawa interactions, just like the Higgs.
Clearly, many models are possible.

A predictive model with no extra parameters is obtained by introducing a second copy of the SM and by imposing a Z$_2$ symmetry,
spontaneously broken by the fact that the mirror Higgs field, identified with $S$, lies in the Planck-scale
minimum while the Higgs field  lies in the weak scale minimum.
The mirror SM photon would be massless.
Depending on the thermal history of the universe, a heavy mirror SM particle, such as a mirror neutrino or electron, 
 could be a Dark Matter candidate. The interactions between such dark matter candidate with the visible sector are suppressed by $\lambda_{HS}$; as we will discuss in section \ref{weak-scale-gem}, the smallness of $\lambda_{HS}$ is implied by a  mechanism to understand the hierarchy between the Planck and the electroweak scales.
%\footnote{\xxx{.....XXX..}The mirror QCD scale would
%be... the mirror lightest baryon would be $uuu$ with mass; the mirror neutrino mass cannot be computed
%without knowing how normal neutrinos get a mass, but could be $m'_\nu \sim m_\nu M_{\rm Pl}/M_h\sim 100\TeV$.
%The mirror neutrino is a Dark Matter candidate.}

\section{Inflation}\label{infl}
Inflation with a small amplitude of perturbations is not a typical outcome of quantum field theory: it needs potentials with special flatness properties
and often super-Planckian vacuum expectation values.
Agravity allows to compute the effective action at super-Planckian vacuum expectation values,
and potentials are flat at tree level, when expressed in the Einstein frame.  
At loop level quantum corrections lead to deviations from flatness, 
proportional to the $\beta$ functions of the theory computed in section~\ref{RGE}:
thereby perturbative couplings lead to  quasi-flat potentials, as suggested by inflation.

\medskip

All scalar fields in agravity are inflaton candidates: 
the Higgs boson $h$, the Higgs of gravity $s$,
the scalar component of the graviton $\chi$.
In order to make $\chi$ explicit, we eliminate the $R^2/6\gs^2$ term in the Lagrangian by adding an auxiliary field $\chi$ 
and its vanishing  action
$-\sqrt{|\det g}|  {(R+3 f_0^2 \chi/2)^2}/{6f_0^2}$.  Next, by performing a Weyl rescaling\footnote{The agravity Lagrangian
of eq.\eq{Ladim}
also contains the square of the Weyl tensor,
$(\frac13 R^2 -  R_{\mu\nu}^2)/{\gt^2} $,
that remains unaffected and does not contribute to classical cosmological evolution equations.
A similar procedure allows to make also the massive spin-2 state explicit~\cite{Ovrut}.
}
$g_{\mu\nu}^E = g_{\mu\nu}\times f/\bar M_{\rm Pl}^2$
with $f = \xi_S s^2+\xi_H h^2+\chi$ one obtains a canonical Einstein-Hilbert term,
mixed kinetic terms for the scalars,
\beq \Lag =
 \sqrt{\det g_E} \bigg[   - \frac{\bar M_{\rm Pl}^2}{2} R_E
+  \bar M_{\rm Pl}^2 \bigg( \frac{(\partial_\mu s)^2+(\partial_\mu h)^2}{2f} 
 +  \frac{3 (\partial_\mu f)^2}{4 f^2}\bigg)
 - V_E
  \bigg]+\cdots\eeq
as well as their effective potential:
\beq V_E = \frac{\bar M_{\rm Pl}^4}{f^2}\left(V(h,s) + \frac{3\gs^2}{8} \chi^2\right). \eeq
Disentangling the kinetic mixing between $s$ and $\chi$ around the minimum of $V_E$, we find mass eigenstates
\beq\label{eq:mm}
M_\pm^2 =\frac{m_s^2+m_\chi^2}{2}\pm \frac{1}{2}\sqrt{(m_s^2+m_\chi^2)^2-4\frac{m_s^2m_\chi^2}{1+6\xi_S}}
% \frac{ \langle s\rangle^2}{4} \left[ f_0^2 \left( 1+6 \xi_S \right) \xi_S+\frac{b}{2} \pm\sqrt{f_0^4\xi_S^2(1+6\xi_S)^2+b\xi_S f_0^2 (6\xi_S-1)+\frac{b^2}{4}}\right],
\eeq  
where $m_s^2\equiv \langle s\rangle^2 b /4$, $m_\chi^2\equiv \langle s\rangle^2 f_0^2 (1+6\xi_S)\xi_S/2$ and  
$b=\beta(\beta_{\lambda_S})$  is the $\beta$ function of the $\beta$ function  of $\lambda_S$ evaluated at $\langle s\rangle$.

Various regimes for inflation are possible.  
In the limit where $h$ or $\chi$ feel the vacuum expectation value of $s$ as a constant mass term,
one obtains Starobinsky inflation and Higgs $\xi$-inflation~\cite{xii}: agravity dictates how they can hold above the Planck scale.
Leaving a full analysis to a future work, we here want to explore the possibility that the inflaton is the field $s$
that dynamically generates the Planck scale, as discussed in section~\ref{PlDyn}.

In the limit where the spin 0 graviton $\chi$ is heavy enough that we can ignore its kinetic mixing with $s$,
we can easily convert $s$ into a scalar $s_E$ with
canonical kinetic term, as discussed in eq.\eq{LsE}.
Then, the usual formalism of slow-roll parameters allows to obtain the inflationary predictions.
The slow-roll parameters $\epsilon$ and $\eta$ are given by the $\beta$ functions of the theory.  At leading order we find:
\begin{eqnsystem}{sys:slowroll}
 \epsilon  &\equiv & \frac{\bar M_{\rm Pl}^2}{2} \left(\frac{1}{V_E} \frac{\partial V_E}{\partial s_E}\right)^2=
\frac12 \frac{\xi_S}{1+6\xi_S} \bigg[ \frac{\beta_{\lambda_S}}{\lambda_S} - 2 \frac{\beta_{\xi_S}}{\xi_S}\bigg]^2  ,\\
%{\frac{( {\beta_{\lambda_S}}/{\lambda_S} - 2 {\beta_{\xi_S}}/{\xi_S})^2}{2(6\xi_S-1)/\xi_S -12 \beta_{\xi_S}/\xi_S -3 \beta^2_{\xi_S}/\xi_S^2}}   ,\\
\eta &\equiv& \bar M_{\rm Pl}^2 \frac{1}{V_E} \frac{\partial^2 V_E}{\partial s_E^2} =
\frac{\xi_S}{1+6\xi_S}\bigg[\frac{\beta(\beta_{\lambda_S})}{\lambda_S} - 2\frac{\beta(\beta_{\xi_S})}{\xi_S}
+\frac{5+36\xi_S}{1+6\xi_S}\frac{ \beta_{\xi_S}^2}{\xi_S^2}-\frac{7+48\xi_S}{1+6\xi_S} \frac{\beta_{\lambda_S}\beta_{\xi_S}}{2\lambda_S\xi_S}
\bigg].~~~~
\end{eqnsystem}
The  scalar amplitude $A_s$, its  spectral index $n_s$ and the tensor-to-scalar ratio $r = A_t/A_s$ are predicted as
\beq n_s = 1 - 6\epsilon + 2\eta,\qquad A_s = \frac{V_E/\epsilon}{24\pi^2 \bar M_{\rm Pl}^4},\qquad
 r = 16\epsilon \eeq
where all quantities are evaluated at about $N\approx 60$ $e$-folds before the end of inflation,
when the inflation field $s_E(N)$ was  given by
\beq N = \frac{1}{\bar M_{\rm Pl}^2} \int_0^{s_E(N)} \frac{V_E(s_E)}{V'_E(s_E)} ds_E.\eeq
In any given agravity model the running of $\lambda_S$ and of $\xi_S$
and consequently the inflationary predictions can be computed numerically.

\medskip

\subsection{Agravity inflation: analytic approximation}\label{iapprox}
We consider a simple analytic approximation
that encodes the main features of this scenario.
As discussed in section~\ref{PlDyn} and summarised in eq.\eq{agravMPl},
dynamical generation of the Planck scale with vanishing cosmological constant demands
that the quartic $\lambda_S$ as well as its $\beta$ function vanish at a scale $\langle s\rangle=\bar M_{\rm Pl}/\sqrt{\xi_S}$.  Thereby, around such minimum, we can approximate the running parameters as
\beq \label{eq:quadratic}
\lambda_S(\mub\approx s) \approx \frac{b}{2} \ln^2 \frac{s}{\langle s\rangle},\qquad \xi_S(\mub) \approx \xi_S. \label{run-parameters}\eeq
We neglect the running of $\xi_S$, given that it does not need to exhibit special features.
The coefficient $b=\beta(\beta_{\lambda_S})$ can be rewritten as $b\equiv {g^4}/{(4\pi)^4}$,
where $g^4$ is the sum of quartic powers of the adimensional couplings of the theory.
It can be computed in any given model. 
%\footnote{Within the 
%
%When the potential is $\lambda_S s^4/4$ where $\lambda_S$ is given by (\ref{run-parameters}) the mass spectrum   is 
%\bea 
%M_{\rm gr}^2&=&0, \quad M_2^2 = \frac12 \gt^2 \xi_S \langle s\rangle^2, \\ 
%  M_{\pm}^2&=&\frac{ \langle s\rangle^2}{4} \left[ f_0^2 \left( 1+6 \xi_S \right) \xi_S+b \pm\sqrt{f_0^4\xi_S^2(1+6\xi_S)^2+2b\xi f_0^2 (6\xi_S-1)+b^2}\right],
%\eea
%where $b\equiv g^4/2(4\pi)^4$. $M_{\rm gr}^2=0$ confirms that the graviton remains massless, $M_2$ is the mass of the other spin-2 state and $M_\pm$ the masses of the scalar states (the result of the mixing between the one coming from $h_{\mu\nu}$ and that from $s$).
%
%In the limit of $M_0\gg M_s$ one recovers the formula for $M_s$.
%
%}
With this approximated running the slow-roll parameters of eq.s~(\ref{sys:slowroll}) simplify to
\beq \epsilon  \approx \eta \approx \frac{2\xi_S}{1+6\xi_S}\frac{1}{\ln^2 s/\langle s\rangle} = \frac{2\bar M_{\rm Pl}^2}{s_E^2}
.\eeq
The latter equality holds because, within the assumed approximation, the explicit expression for the Einstein-frame scalar $s_E$ is
\beq s_E = \bar M_{\rm Pl}\sqrt{\frac{1+6\xi_S}{\xi_S}}\ln \frac{s}{\langle s\rangle}.\eeq
The Einstein-frame potential gets approximated, around its minimum, as a quadratic potential:
\beq \label{eq:m}
V_E = \frac{\bar M_{\rm Pl}^4}{4} \frac{\lambda_S}{\xi_S^2 }\approx \frac{M_s^2}{2} s_E^2\qquad\hbox{with}\qquad
M_s = \frac{g^2\bar M_{\rm Pl} }{2(4\pi)^2} \frac{1}{\sqrt{\xi_S(1+6\xi_S)}}.\eeq
Notice that the eigenvalue $M_-$ of eq.\eq{mm} indeed reduces to $M_s$, in the limit where it is much lighter than the other eigenvalue
$M_0^2 \simeq \frac12 \gs^2\bar M_{\rm Pl}^2 (1+6\xi_S)$.

%so that eq.s~(\ref{sys:slowroll}) reduce to $\epsilon \approx \eta \approx 2\bar M_{\rm Pl}^2/s_E^2$
%with 
Inserting 
the value of $s_E$ at $N\approx 60$ $e$-folds before the end of inflation, $s_E(N) \approx 2\sqrt{N}\bar M_{\rm Pl}$,
we obtain the predictions:
\beq n_s\approx 1-\frac{2}{N}\approx 0.967,\qquad
r\approx \frac{8}{N}\approx 0.13,\qquad
A_s \approx \frac{b N^2}{24 \pi^2 \xi_S (1+6\xi_S)} . \eeq
Such predictions are typical of quadratic potentials, and this is a non-trivial fact.

Indeed, vacuum expectation values above the Planck scale, $s_E \approx 2\sqrt{N} \bar M_{\rm Pl}$,
are needed for inflation from a quadratic potential and, more generically,
if the tensor/scalar ratio is above the Lyth bound~\cite{Lyth}.
This means that, in a generic context, higher order potential terms suppressed by the Planck scale become important,
so that the quadratic approximation does not hold.

Agravity predicts physics above the Planck scale,
and a quadratic potential is a good approximation, even at super-Planckian
vev, because coefficients of higher order terms are dynamically suppressed by extra powers
of the loop expansion parameters, roughly given by $g^2/(4\pi)^2$.
Higher order terms are expected to give corrections of relative order $g^2\sqrt{N}/(4\pi)^2$,
which are small if the theory is weakly coupled.

\subsection{Numerical model-dependent computation}
We here consider the specific model presented in section~\ref{mod},
where the scalar $S$ is identified with the Higgs 
doublet of a mirror sector which is an exact copy of the SM, with the only difference
that $S$ sits in the Planck-scale minimum of the SM effective potential.

This model predicts that the $\beta$ function coefficient in eq.\eq{quadratic}
equals $g^4\approx 1.0$ 
provided that we can neglect the  gravitational couplings $\gs,\gt$ with respect to the
known order-one SM couplings $y_t,g_3,g_2,g_1$.

Thereby the observed scalar amplitude $A_s= 2.2~10^{-9}$~\cite{Planck}
is reproduced for $\xi_S \approx 210$.
A large $\xi_S$ is perturbative as long as it is smaller than $1/f_{0,2}$.

%(As discussed in section~\ref{RGExi}, $\xi'_S$ is the natural parameter that gives a simpler form to the RGEs).

We notice that $\xi_S$ is not a free parameter, within the context of the
SM mirror model: 
the vev of the Higgs mirror $s$ is given by the RGE scale at which $\beta_{\lambda_S}$ vanishes
(see fig.\fig{RGESM}b), and in order to reproduce the correct Planck scale with $\xi_S \approx 210$
one needs  $\langle s\rangle = \bar M_{\rm Pl}/\sqrt{\xi_S} = 1.6~10^{17}\GeV$.
The fact that this condition can be satisfied (within the uncertainties) is a test of the model.

The inflaton mass $M_s\approx 1.4~10^{13}\GeV$  is below the Planck scale because suppressed by the $\beta$-functions of the theory, see eq.\eq{m}.

\medskip

The model allows to compute the full inflationary potential from the full
running of $\lambda_S$ (shown in fig.\fig{RGESM}) 
and of $\xi_S$.
The computation is conveniently performed in the Landau gauge $\xi_V=0$, given that the
gauge-dependence of the effective potential gets canceled by the gauge-dependence of the scalar kinetic term~\cite{Sher}.
By performing a numerical computation 
we find a more precise prediction $r\approx 0.128$ for $N\approx 60$.
This is compatible with the expected accuracy of the quadratic approximation,
estimated as $g^2\sqrt{N}/(4\pi)^2\approx 5\%$ in section~\ref{iapprox}.

%a numerically similar prediction  , which is compatible with present data~\cite{Planck}.
%The vev $\langle s\rangle = \bar M_{\rm Pl}/\sqrt{\xi_S}\approx 10^{17}\GeV$
%is close to the scale at which the SM-mirror $\beta_{\lambda_S}$ vanishes.
%\xxx{presentare come predizione per $A_s$ dato $M_{\rm Pl}$?}

%The observed amount of scalar inhomogeneities $A_s= 2.2~10^{-9}$ is
%reproduced for
%$V_E (s_E(60))/\epsilon \approx (0.0269 \bar M_{\rm Pl})^4$.

\bigskip

In conclusion,
we identified the inflaton with the field that dynamically generates the Planck scale.
In the agravity context, such field must have a dimensionless logarithmic potential: this is why
our predictions for $r\approx 8/N\approx 0.13$ differ from the tentative prediction $r\approx 12/N^2\approx 0.003$ of a generic $\xi$-inflation model with mass parameters
in the potential~\cite{xii}.

\section{Dynamical generation of the Weak scale} \label{weak-scale-gem}

In section~\ref{PlDyn} we discussed how the Planck scale can be dynamically generated.
We now discuss how it is possible to generate also the electro-weak scale,
such that it {\em naturally} is much below the Planck scale.
`Naturally' here refers to the modified version of naturalness adopted in~\cite{FN}, where quadratically divergent
corrections are assumed to vanish, such that no new physics is needed at the weak scale to keep it stable.
The present work proposed a theoretical motivation for the vanishing of power divergences: they have mass dimension, and thereby must vanish if the fundamental theory contains no dimensionful parameters.
This is the principle that motivated our study of adimensional gravity.

In this scenario, the weak scale can be naturally small, and the next step is exploring what can be the
physical dynamical origin of the small ratio $M_h^2/M_{\rm Pl}^2 \sim 10^{-34}$.
The dynamics that generates the weak scale can be:
\begin{itemize}
\item[a)] {\bf around the weak scale}, with physics at much high energy only giving negligible finite corrections to the Higgs mass.
Models of this type have been proposed in the literature~\cite{FNmodels}, although the issue of gravitational corrections has not been addressed. Such models lead to observable signals in weak-scale experiments.

\item[b)] {\bf much above the weak scale}.  For example, Einstein gravity naively suggests that any particle with mass $M$ gives a finite gravitational correction to the Higgs mass
at three~\cite{AD} and two loops:
\beq \delta M_h^2 \sim \frac{y_t^2 M^6}{(4\pi)^6 M_{\rm Pl}^4} + \frac{\xi_H M^6}{(4\pi)^4 M_{\rm Pl}^4}\eeq
which is of the right order of magnitude for $M\sim 10^{14}\GeV$.

\end{itemize}
In the context of agravity we can address the issue of gravitational corrections, and propose a scenario where the weak scale is generated from the Planck scale.
It is convenient to divide the computation into 3 energy ranges

\begin{itemize}

\item[1)] {\em Low energies}: at RGE scales below the mass
$M_{0,2}$ of the heavy gravitons, agravity can be neglected 
and the usual RGE of the SM apply.  The Higgs mass parameter receives
a multiplicative renormalization:
\bea
(4\pi)^2\frac{d  M_h^2 }{d\ln\mub}&=&  {M_h^2}\beta_{M_h}^{\rm SM},\qquad
\beta_{M_h}^{\rm SM}=12 \lambda_H +6  y_t^2 -\frac{9 g_ 2^2}{2} -\frac{9 g_ 1^2 }{10}.
\eea

\item[2)] {\em Intermediate energies} between $M_{0,2}$ and $M_{\rm Pl}$: 
 agravity interactions cannot be neglected but
$M_h$ and $M_{\rm Pl}$ appear in the effective Lagrangian as apparent
dimensionful parameters.
We find that their RGE are gauge-dependent because the unit of mass is gauge dependent.
The RGE for adimensional mass-ratios are  gauge-independent and we find\footnote{We also verified that the RGE for the ratio of
scalar to fermion masses is gauge invariant.
We cannot comparare our eq.\eq{RGEm} with gauge-depend
RGE for $M_{\rm Pl}$ computed in the literature~\cite{Narain,Shapiro,Avramidi,Dono} with discrepant results,
given that we use a different gauge. }
\bea
(4\pi)^2\frac{d  }{d\ln\mub} \frac{M_h^2}{\bar M_{\rm Pl}^2}&=& -\xi_H [5\gt^4+\gs^4(1+6\xi_H)]
-\frac13 \bigg( \frac{M_h^2}{\bar M_{\rm Pl}^2}\bigg)^2(1+6\xi_H)+\nonumber\\
&& + \frac{M_h^2}{\bar M_{\rm Pl}^2} 
\bigg[\beta_{M_h}^{\rm SM}+5\gt^2 +\frac{5}{3}\frac{\gt^4}{\gs^2} + \gs^2 (\frac13 + 6\xi_H + 6 \xi_H^2)\bigg].\label{eq:RGEm}
\eea
The first term is crucial: it describes corrections to $M_h$ proportional to $M_{\rm Pl}$.
A naturally small~\cite{FN} weak scale arises provided that the agravity couplings are small:
\beq \gs,\gt \approx \sqrt{\frac{4\pi M_h}{M_{\rm Pl}}}\sim 10^{-8}.\eeq
The mass of the spin-2 graviton ghost is
$M_2 = \gt \bar M_{\rm Pl}/\sqrt{2}\approx  3~10^{10}\GeV$.
The spin-0 massive component of the graviton mixes with the other scalars giving rise to
the mass eigenvalues of eq.\eq{mm}.
Experimental bounds are safely satisfied.

\item[3)] {\em Large energies} above the Planck mass: the theory is adimensional and the RGE of section~\ref{RGE} apply.
According to the Lagrangian of eq.\eq{LadimBSM}, 
the quartic coupling $\lambda_{HS} |H|^2 |S|^2$
leads to a Higgs mass term $\frac12 M_h^2 |H|^2$ given by
 $M_h^2 = \lambda_{HS} \langle s\rangle^2$.
  Ignoring gravity, $\lambda_{HS}$ can be naturally arbitrarily small, because it is the only interaction
 that couples the SM sector with the $S$ sector.
Within agravity, a non vanishing $\lambda_{HS}$ is unavoidably generated by RGE running at one-loop order, 
as shown by its RGE in eq.\eq{RGElambdaHS}, which contains the non-multiplicative contribution:
\beq (4\pi)^2 \frac{d{\lambda_{HS}}}{d\ln\mub} = -\xi_H\xi_S [ 5\gt^4  + \gs^4(6\xi_S+1)(6\xi_H+1)] + \cdots .\eeq
For $\xi_S=0$ this equation is equivalent to\eq{RGEm}.
We need to assume that the mixed quartic acquires its minimal natural value,
$\lambda_{HS}\sim f_{0,2}^4$
(for simplicity we do not consider the possibility of  values of $\xi_{H,S}=\{0,-1/6\}$ that lead to special cancellations).
\end{itemize}
In conclusion, agravity unavoidably generates a contribution to the Higgs mass given by
\beq  M_h^2 \approx
\frac{\bar M_{\rm Pl}^2  \xi_H}{(4\pi)^2}  [ 5\gt^4   + \gs^4(6\xi_S+1)(6\xi_H+1)] \ell  \eeq
where $\ell$ is a positive logarithmic factor.

This alternative understanding of the Higgs mass hierarchy problem relies on the smallness of some parameters.
All parameters assumed to be small are naturally small, 
just like the  Yukawa coupling of the electron, $y_e\sim 10^{-6}$, is naturally small.
These small parameters do not receive unnaturally large quantum corrections.  No fine-tuned cancellations are necessary.

At perturbative level, this is clear from the explicit form of the one-loop RGE equations derived in section~\ref{RGE}:
quantum corrections to $\gs,\gt$ are proportional to cubic powers of $\gs,\gt$,
and higher order loop corrections are even more suppressed.

\smallskip

At non perturbative level,  a black hole of mass $M$ might give a quantum correction of
order $\delta M_h^2 \sim M^2_{\rm BH} e^{-S_{\rm BH}}$ where $S_{\rm BH} = M^2_{\rm BH}/2\bar M_{\rm Pl}^2$ is the black hole entropy.
Black holes with Planck-scale mass might give an unnaturally large correction, $\delta M_h^2 \gg M_h^2$,
ruining naturalness.  Planck-scale black holes do not exist in agravity, where the minimal mass of a black hole
is $M_{\rm BH} \circa{>} \bar M_{\rm Pl}/f_{0,2}$, as clear from the fact that the massive anti-gravitons
damp the $1/r$ Newton behaviour of the
gravitational potential at $r\circa{<}1/M_{0,2}$:
\beq V%=  h^{00} 
  =- \frac{Gm}{r} \left[1- \frac43 e^{-M_2 r}  + \frac13 e^{-M_0 r} \right].\eeq
Thereby, non-perturbative quantum corrections are expected to be negligible in agravity,
because exponentially suppressed as $e^{-1/f_{0,2}^2}$.

\bigskip

%The quantum behaviour of the theory splits in 3 regimes:
%\begin{itemize}
%\item[a)] at energies much above the Planck mass the theory is adimensional and the RGE of section~\ref{RGE} apply;
%\item[b)] at energies between $M_{0,2}$ and $M_{\rm Pl}$ the agravity terms must be included in the RGE.
%Furthermore, the massive parameters $M_h$ and $M_{\rm Pl}$ appear in the effective Lagrangian.
%We find that their RGE are gauge-dependent because the unit of mass is gauge dependent.\footnote{We cannot compare with previous partial computations given that we use a different gauge~\cite{Narain,Shapiro,Avramidi,Dono}.}
%The RGE for adimensional mass-ratios are  gauge-independent and we find
%\bea
%(4\pi)^2\frac{d  }{d\ln\mub} \frac{M_h^2}{\bar M_{\rm Pl}^2}&=& -\xi_H [5\gt^4+\gs^4(1+6\xi_H)]
%+ \frac{M_h^2}{\bar M_{\rm Pl}^2} 
%\bigg[12 \lambda_H +6  y_t^2 -\frac{9 g_ 2^2}{2} -\frac{9 g_ 1^2 }{10}+ \nonumber\\
%&& +5\gt^2+\frac{5}{3}\frac{\gt^4}{\gs^2} + \gs^2 (\frac13 + 6\xi_H + 6 \xi_H^2)\bigg]
%+ ( \frac{M_h^2}{\bar M_{\rm Pl}^2})^2(1+6\xi_H)
%\eea
%\xxx{fattori overall a caso}

% SECONDO NARAIN
%\beq
%(4\pi)^2\frac{d M_{\rm Pl}^2 }{d\ln\mub} =M_{\rm Pl}^2 \bigg[\frac53  \frac{g_2^4}{g_0^2}-\frac{7}{24} g_0^2\bigg].
%\eeq
%  SECONDO AVRAMIDI
%\beq
%%(4\pi)^2\frac{d M_{\rm Pl}^2 }{d\ln\mub} =M_{\rm Pl}^2 \bigg[-\frac53  \frac{g_2^4}{g_0^2}+\frac{13}{6}\gt^2+\frac12 \gs^2\bigg].
%%\eeq
%\item[c)] at energies below $M_{0,2}$ the usual RGE of the SM apply.
%\end{itemize}

The Higgs of gravity $s$ has a mass $M_s$ which can be anywhere between the weak scale and the Planck scale,
depending on how large are the gauge and Yukawa couplings within its sector.
Its couplings to SM particles are always negligibly small.
In the model where $s$ is the Higgs of a mirror copy of the SM, its mass is a few orders of magnitude below the Planck scale.

As a final comment, 
we notice that accidental global symmetries (a key ingredient of axion models)
are a natural consequence of the dimensionless principle.
In the usual scenario, ad hoc model building is needed in order to suppress
explicit breaking due to mass terms or non-renormalizable operators~\cite{Randall}.
An axion can be added to agravity compatibly with finite naturalness along the lines of~\cite{FN}.

\section{Conclusions}
In conclusion, we proposed that the fundamental theory contains no dimensionful parameter.  
Adimensional gravity (agravity for short) is renormalizable because gravitons have a kinetic term with 4 derivatives and two adimensional coupling constants $\gs$ and $\gt$.

The theory predicts physics above the Planck scale. We computed the RGE of a generic agravity theory,
see eq.s\eq{RGEY},\eq{RGElambda},\eq{RGExi} and (\ref{sys:RGG}).
We found that quantum corrections can dynamically generate the Planck scale as the vacuum expectation value of a scalar $s$,
that acts as the Higgs of gravity.
The cosmological constant can be tuned to zero.
This happens when a running quartic coupling and its $\beta$ function both vanish around the Planck scale, as summarised in eq.\eq{agravMPl}.
The quartic coupling of the Higgs in the SM can run in such a way, see fig.\fig{RGESM}.

The graviton splits into the usual massless graviton, into a massive spin 2 anti-graviton, and into a scalar.
The spin 2 state is a ghost, to be quantised as a state with positive kinetic energy but negative norm.

The lack of dimensional parameters implies successful quasi-flat inflationary potentials at super-Planckian 
vacuum expectation values:  the slow-roll parameters are the $\beta$ functions of the theory. 
Identifying the inflaton with the Higgs of gravity leads to predictions $n_s\approx 0.967$ for the spectral index 
and $r\approx 0.13$ for the tensor/scalar amplitude ratio.

The Higgs of gravity can also be identified with the Higgs of the Higgs:
if $\gs,\gt\sim 10^{-8}$ are small enough, gravitational loops generate the observed weak scale.
In this context, a  weak scale much smaller than the Planck scale is natural: all small parameters receive small quantum corrections. 
In particular, quadratic divergences must vanish in view of the lack of any fundamental dimensionful parameter,
circumventing the usual hierarchy problem.

\small

\subsubsection*{Acknowledgments}
We thank P. Benincasa, S. Dimopoulos, J. Garcia-Bellido, G. Giudice, J. March-Russell, P. Menotti, M. Redi, M. Shaposhnikov for useful discussions.
This work was supported by the {\sc ESF} grant MTT8 and by the SF0690030s09 project
and by the European Programme PITN-GA-2009-237920 (UNILHC). We are grateful to CERN for hospitality. 
The work of Alberto Salvio has been also supported by the Spanish Ministry of Economy and Competitiveness under grant FPA2012-32828, Consolider-CPAN (CSD2007-00042), the grant  SEV-2012-0249 of the ``Centro de Excelencia Severo Ochoa'' Programme and the grant  HEPHACOS-S2009/ESP1473 from the C.A. de Madrid.

\end{document}